\documentclass[aps,pre,showpacs,showkeys,amssymb,twocolumn]{revtex4}
\usepackage{bm}
\usepackage{graphicx}
\usepackage{amsfonts}
\usepackage{amsmath}
\usepackage{amssymb}

\begin{document}

\title{Topology of configuration space of the mean-field $\phi^4$ model by Morse theory}

\author{Fabrizio Baroni}

\email{f.baroni@ifac.cnr.it, baronifab@libero.it}
\affiliation{CNR - Istituto di fisica applicata Nello Carrara, Sesto Fiorentino (FI), Italy}

\date{\today}

\begin{abstract}
In this paper we present the study of the topology of the equipotential hypersurfaces of configuration space of the mean-field $\phi^4$ model with a $\mathbb{Z}_2$ symmetry. Our purpose is discovering, if any, the relation between the second-order $\mathbb{Z}_2$-symmetry breaking phase transition and the geometric entities mentioned above. The mean-field interaction allows us to solve analytically either the thermodynamic in the canonical ensemble or the topology by means of Morse theory. We have analyzed the results at the light of two theorems on sufficiency conditions for symmetry breaking phase transitions recently proven. This study makes part of a research line based on the general framework of geometric-topological approach to Hamiltonian chaos and critical phenomena.
\end{abstract}



\maketitle

%{\setlength\arraycolsep{2pt}

\section{Introduction}

Phase transitions are sudden changes of the macroscopic behaviour of a physical system composed by many interacting parts occurring while an external parameter is smoothly varied, generally the temperature, but e.g. in a quantum phase transition it is the external magnetic field. From a mathematical viewpoint, a phase transition is a non-analytic point in the partition function emerging as the thermodynamic limit has been performed. The successful description of phase transitions starting from the properties of the microscopic interactions among the components of the system is one of the major achievements of equilibrium statistical mechanics.

From a statistical-mechanical point of view, in the canonical ensemble, a phase transition occurs at special values of the temperature $T$ called transition points, where thermodynamic quantities such as pressure, magnetization, or heat capacity, are non-analytic functions of $T$. These points are the boundaries between different phases of the system. Starting from the exact solution of the $2$-dimensional Ising model \cite{ising} by Onsager \cite{onsager}, these singularities have been found in many other models, and later developments like the renormalization group theory \cite{goldenfeld} have considerably deepened our knowledge of the properties of the transition points. Typically, but non necessarily,  these singularities are associated with spontaneous symmetry breaking phenomenon, giving rise to symmetry breaking phase transitions (SBPT hereafter). In this paper we consider this case only. But in spite of the success of equilibrium statistical mechanics, the issue of the origin of SBPTs remains open, and this motivates further studies of SBPTs. Solving this issue means finding the characteristics of a system capable of generating a SBPT. 

For a Hamiltonian system it is natural to think that the origin of a SBPT is hidden in the potential, therefore the issue becomes which properties of the potential are \emph{sufficiency conditions} to entail a SBPT. 
The topological-geometrical approach to PTs has been proposed and developed in \cite{ccp,ccp1,ccp2}. For a complete review we refer the reader to \cite{ccp1,k,pettini}, here we will limit to give only the most basic ideas necessary to frame the work of this paper. This approach stems from a very intuitive idea. Consider the canonical treatment of a system. In the thermodynamic limit the representative point is forced to live on a single equipotential surface because the statistical measure shrinks around it. The potential level of this surface is the thermodynamic average potential. For example, if we imagine that this surfaces is made by two disjointed connected components, which are one the image of the other under a Z2 symmetry, then the representative point cannot jump from side to side, so that the symmetry could be broken. 

In \cite{bc} some sufficiency conditions have been established for $\mathbb{Z}_2$-SBPT in systems with long-range interaction condensed in two straightforward theorems. These conditions involve some topological features of the equipotential hypersurfaces and
has already been found at work in various toy models \cite{b4,b,b1}, e.g. the hypercubic one in \cite{bc}. The mean-field $\phi^4$ model, i.e. with all-to-all interaction, satisfies partially the aforementioned conditions, and it is mainly in this light that the results of this paper will be analyzed. 
But before going into the details of the discussion, we define the framework that will be used in the following.

Consider an $N$ degrees of freedom system with Hamiltonian given by
\begin{equation}
H(\textbf{p},\textbf{q})=T+V=\sum_{i=1}^N \frac{p_i^2}{2}+V(\textbf{q}).
\end{equation}
Let $M\subseteq\mathbb{R}^N$ be the configuration space. The partition function is by definition
\begin{eqnarray}
Z(\beta,N)=\int_{\mathbb{R}^N\times M} \rm d\mathbf{p}\,\rm d\mathbf{q}\,e^{-\beta H(\mathbf{p},\mathbf{q})}=\nonumber
\\
=\int_{\mathbb{R}^N} \rm d\mathbf{p}\,e^{-\beta\sum_{i=1}^N \frac{p_i^2}{2}}\int_M \rm d\mathbf{q}\,e^{-\beta V(\textbf{q})}=Z_{kin}Z_c
\end{eqnarray}
where $\beta=1/T$ (in unit $k_B=1$), $Z_{kin}$ is the kinetic part of $Z$, and $Z_c$ is the configurational part. In order to develop what follows we assume the potential to be lower bounded.
$Z_c$ can be written as follows
\begin{eqnarray}
Z_c=N\int_{v_{min}}^{+\infty}\rm dv\,e^{-\beta Nv}\int_{\Sigma_{v,N}}\frac{\rm d\Sigma}{\left\|\nabla V\right\|}
\label{zc}
\end{eqnarray}
where $v=V/N$ is the potential per degree of freedom, and the $\Sigma_{v,N}$'s are the $v$-level sets defined as
\begin{equation}
\Sigma_{v,N}=\{\textbf{q}\in M: v(\textbf{q})=v\}.
\label{sigmav}
\end{equation}
The $\Sigma_{v,N}$'s are the boundaries of the $M_{v,N}$'s ($\Sigma_{v,N}=\partial M_{v,N}$) defined as 
\begin{equation}
M_{v,N}=\{\textbf{q}\in M: v(\textbf{q})\leq v\}.
\label{Mv}
\end{equation}
The set of the $\Sigma_{v,N}$'s is a foliation of configuration space $M$ while varying $v$ between $v_{min}$ and $+\infty$. The $\Sigma_{v,N}$'s are very important submanifolds of $M$ because as $N\rightarrow\infty$ the canonical statistic measure shrinks around $\Sigma_{\bar{v}(T),N}$, where $\bar{v}(T)$ is the average potential per degree of freedom. Thus, $\Sigma_{\bar{v}(T),N}$ becomes the most probably accessible $v$-level set by the representative point of the system.

This fact may have significant consequences on the symmetries of the system and on the analyticity of $Z_c$ because of the non-trivial topology in general of the $\Sigma_{\bar{v}(T),N}$ which changes while varying $T$.

We can make the same considerations for $Z_{kin}$, but the related submanifolds $\Sigma_{t,N}$, where $t=T/N$ is the kinetic energy per degree of freedom, are all trivially homeomorphic to an $N$-sphere, thus they cannot affect the symmetry properties of the system by topological reasons. Furthermore, $Z_{kin}$ is analytic at any $T$ in the thermodynamic limit, so that it cannot entail any loss of analyticity in $Z$.

Let's go back to the results in \cite{bc}.
Leaving aside the details, a sufficiency condition to generate a $\mathbb{Z}_2$-SBPT in a long-range interacting system is that the potential is double-well-shaped with the gap between the two global minima proportional to $N$.
If this condition is satisfied, there exists $v'$ such that the $\Sigma_{v,N}$'s are made by two disjointed connected components, which are one the image of the other under the $\mathbb{Z}_2$ symmetry, for $v\in [v_{min},v')$ where $v_{min}$ is the global minimum. This fact, which involves the topology of the $\Sigma_{v,N}$'s although in a trivial way, entails the $\mathbb{Z}_2$-SB. 
Moreover, if for $v\ge v’$ the $\Sigma_{v,N}$'s are diffeomorphic to a single connected component on which the ergodic hypothesis holds, then the $\mathbb{Z}_2$ symmetry is restored and the system undergoes also a PT with the singularity at least in the magnetization.

The results in this paper show that the mean-field $\phi^4$ model is just provided with a double-well potential with the peculiarities explained above, while the potential of the $\phi^4$ model without interaction, i.e. with no $\mathbb{Z}_2$-SBPT, has no such a feature. 
For precision, the potential of the mean-field $\phi^4$ model is not just a simple double-well, indeed if it were the case it would have only three critical points: two global minima and a saddle. Anyway, the presence of local minima, which give rise to the great complication in the topology of the $\Sigma_{v,N}$'s, is irrelevant for the generating-mechanism of a $\mathbb{Z}_2$-SBPT stated in \cite{bc}.

In Sec. \ref{phi4} we present a detailed analytical study of the canonical thermodynamic and of the topological changes which occur in the manifolds $M_{v,N}$ of the mean-field $\phi^4$ model, which allows a complete and constructive analytical characterization of the topology of the $M_{v,N}$'s and also a computation of their Euler characteristic. In Sec. \ref{phi4noint} we present the same study for a case where no $\mathbb{Z}_2$-SBPT is present, i.e. the $\phi^4$ model without interaction, in order to compare the two cases and obtain hints towards a general understanding of the general relation between topology changes and SBPTs. In both cases we use Morse theory as a mathematical tool that allows one to study the topology of a manifold $M$ in terms of the analytical properties of suitable functions (called Morse functions) $f: M\rightarrow\mathbb{R}$. The connection between this technique and physics is made by choosing the potential per degree of freedom $v=V/N$ as Morse function. In Sec. \ref{results} we try to discover any possible relation between topology and geometry and the $\mathbb{Z}_2$-SBPT of the mean-field $\phi^4$ model at the light of the results obtained in  \cite{b3,bc,ccp,ccp1,ccp2,hk,km,mhk}.

\section{Mean-field $\phi^4$ model}
\label{phi4}

The lattice $\phi^4$ models are a class with an $O(n)$ symmetry for $n\geq 1$. We have restricted our study to the $\phi^4$ model with an $O(1)$ symmetry (known even as $\mathbb{Z}_2$ symmetry) and with mean-field (m-f hereafter) interactions, i.e. every degree of freedom interacts with every other. The Hamiltonian is as follows
\begin{equation}
    H=T+V=\sum_{i=1}^N \left(\frac{\pi^2_i}{2}-\frac{\phi_i^2}{2}+\frac{\phi_i^4}{4}\right)-\frac{J}{2N}\left(\sum_{i=1}^N \phi_i\right)^2.
		\label{V}
\end{equation}
The $\pi_i$'s are the canonically conjugated momenta of the coordinates $\phi_i$'s, $J>0$ is the coupling constant, and $N$ is the number of degrees of freedom.

\subsection{Canonical thermodynamic}

In what follows we will disregard the kinetic terms $\frac{\pi^2_i}{2}$ for the reasons already exposed in the previous Section. The configurational partition function is 
\begin{equation}
Z_c=\int\,d^N\phi\,e^{-\beta\left(\sum^{i=1}_{N}V(\phi_i)-\frac{J}{2N}\left(\sum_{i=1}^N \phi_i\right)^2\right)},
\end{equation}
where
\begin{equation}
V(\phi)=-\frac{\phi^2}{2}+\frac{\phi^4}{4}
\label{vlocal}
\end{equation}
is the local potential. The order parameter, i.e. the magnetization in our case, is
\begin{equation}
m=\frac{1}{N}\sum_{i=1}^N\phi_i,
\label{m}
\end{equation}
which, introduced in $Z_c$, gives
\begin{equation}
Z_c=\int d^N\phi\,e^{-\beta\left(\sum_{i=1}^{N}V(\phi_i)-\frac{JN}{2} m^2\right)}, 
\label{Zc}
\end{equation}

Now, for the sake of completeness, we briefly recall the solution of the thermodynamic by means of m-f theory, but it is available in literature, for example in \cite{dl}. M-f interactions imply that the potential is a function of $m$, so that we can analytically solve $Z_c$ by the Hubbard-Stratonovich transformation \cite{goldenfeld} based on the equality

\begin{equation}
    e^{\mu m^2}=\frac{1}{\sqrt{\pi}}\int dy\,e^{-y^2+2\sqrt{\mu}my},
\end{equation}
which, inserted in (\ref{Zc}), yields
\begin{equation}
    Z_c=\frac{1}{\sqrt{\pi}}\int dy\left(\int d\phi\,e^{-\beta V(\phi)+\sqrt{\frac{2\beta J}{N}}m\phi}\right)^N e^{-y^2}.
\end{equation}
After introducing
\begin{equation}
    \varphi(m,\beta)=\ln\int dq\,e^{-\beta \left(V(q)+J m q\right)},
\end{equation}
and the variable changing $y=\sqrt{\frac{N\beta J}{2}}m$, we get
\begin{equation}
    Z_c=\sqrt{\frac{N\beta J}{2\pi}}\int dm\,e^{-N\beta f_c(m,\beta)},
\end{equation}
where
\begin{equation}
    f_c=-\frac{J}{2}m^2+\frac{1}{\beta}\varphi (m,\beta)
\end{equation}
is the configurational free energy per degree of freedom. 

\begin{figure}
\begin{center}
\includegraphics[width=0.235\textwidth]{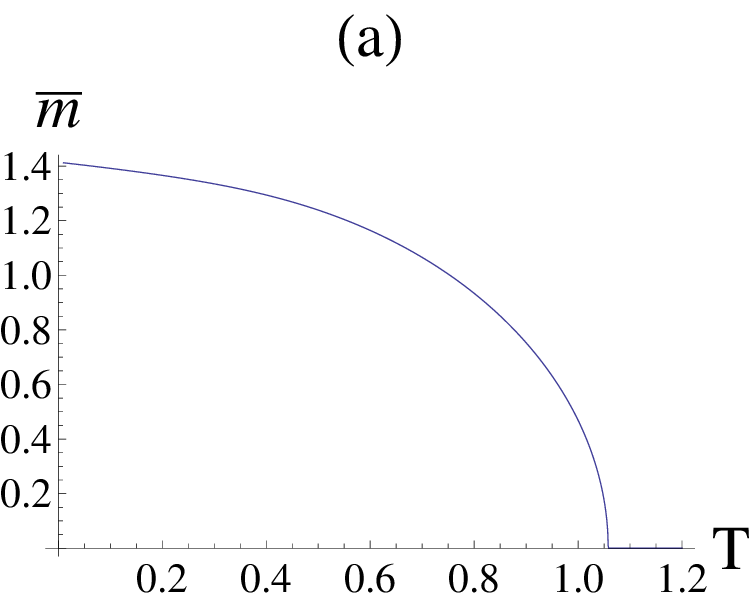}
\includegraphics[width=0.235\textwidth]{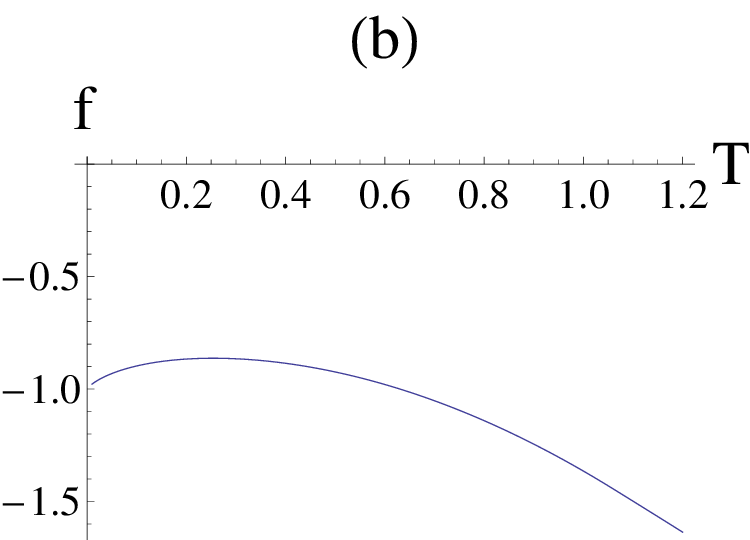}
\includegraphics[width=0.235\textwidth]{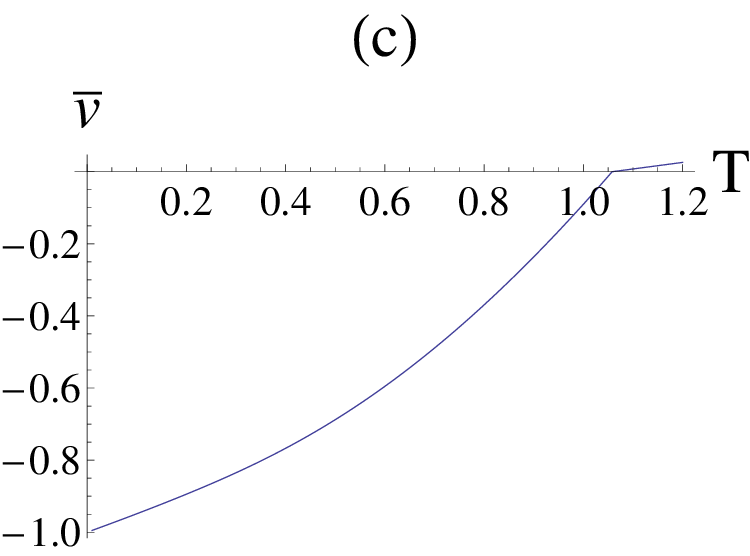}
\includegraphics[width=0.235\textwidth]{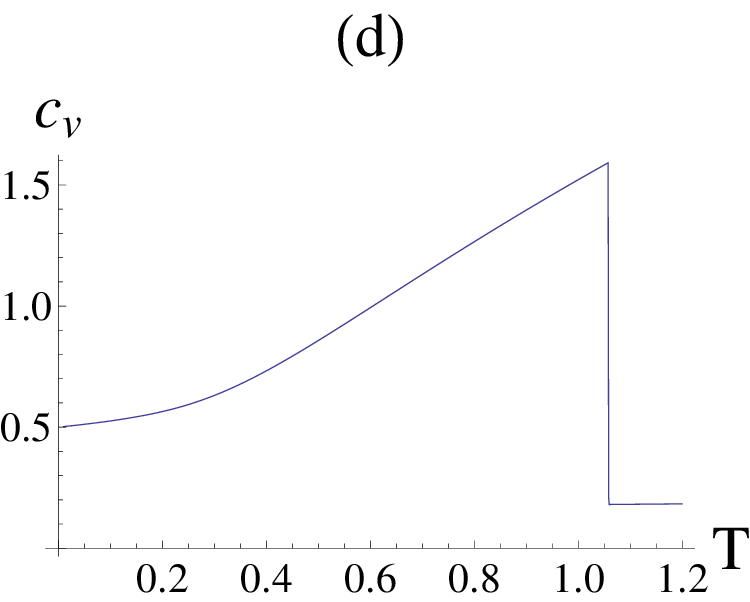}
\caption{Mean-field $\phi^4$ model (\ref{V}) with coupling constant $J=1$. (a), (b), (c), (d) Spontaneous magnetization $\overline{m}$, free energy $f$, specific average potential $\overline{v}$, and specific heat $c_v$, respectively, as functions of the temperature $T$.}
\label{phi4_thermo}
\end{center}
\end{figure}

Finally, in order to apply the saddle point evaluation to calculate $Z_c$ in the thermodynamic limit, we minimize $f_c$ with respect to $m$ at fixed $\beta$ obtaining the spontaneous magnetization $\overline{m}(\beta)$. From the latter we get the free energy, the average potential, and the specific heat
\begin{eqnarray}
    f_c(\beta)=f_c(\overline{m}),
		\\
		\overline{v}(\beta)=-\frac{\partial\varphi}{\partial\beta}+\frac{J}{2}\overline{m}^2,
		\\
		c_v(\beta)=\frac{d\overline{v}}{dT},
\end{eqnarray}
respectively. They are plotted in Fig. \ref{phi4_thermo} as functions of $T$. The picture is the well known one of a second-order $\mathbb{Z}_2$-SBPT with classical critical exponents. 

The critical temperature $T_c$ can been evaluated numerically by the graph of the magnetization with an arbitrary precision. $T_c$ is an increasing function of $J$, the parabola $0.61 J+0.44 J^2$ is a good fit at least up to $J=5$. The critical potential $\overline{v}_c$ is nothing but the value of the average potential at $T_c$. The left panel of Fig. \ref{v'} shows it as a function of $J$.

\subsection{Topology of the submanifolds $M_{v,N}$'s by Morse theory}
\label{phi4top}

Morse theory allows us to characterize the topology of the submanifolds $M_{v,N}$ of configuration space $M=\mathbb{R}^N$ defined in (\ref{Mv}) by a Morse function $f: M\rightarrow\mathbb{R}$. The last is a function whose critical points are non-degenerate, i.e. such that the Hessian matrix of $f$ has rank $N$ at any critical point. For some introductory details of Morse theory we refer to App. \ref{Mt}. 

The potential $V: \mathbb{R}^N\rightarrow\mathbb{R}$ is our Morse function. We cannot show in advance that $V$ is a Morse function, but we have verified that the output of our analysis is a set of isolated critical points. The fact that a critical points is isolated does not imply that it is also non-degenerate, but anyway if the set of critical points are isolated, Morse theory con be successfully applied the same, even though the potential $V$ cannot be properly considered a Morse function. For more information about this delicate question we refer to \cite{mhk}.

The fact that all the critical points are isolated is not surprising, because the set of the Morse functions, or non-properly Morse functions in the sense above-specified, is dense in the set of the smooth functions. Furthermore, the discreteness of the $\mathbb{Z}_2$ symmetry does not create the problem created by continuous symmetries which entail sets of critical points describing submanifolds of configuration space. 

Since in the thermodynamic limit the canonical statistical measure shrinks around the $\Sigma_{v,N}$ corresponding to the average potential density $\overline{v}$, we are interested in the topology of the $\Sigma_{v,N}$'s rather than the $M_{v,N}$'s. Anyway, the topology of the $\Sigma_{v,N}$'s are strictly related to that of the $M_{v,N}$'s, in particular if the $M_{v,N}$'s are diffeomorphic in an interval $[a,b]$, then the same holds also for the $\Sigma_{v,N}$'s.

The critical points are the stationary points of $V$, i.e. the solutions of $\nabla V=0$, which for the potential of the Hamiltonian (\ref{V}) takes the form
\begin{equation}
   \phi^3_i-\phi_i-Jm=0\quad\quad i=1,\cdots,N,
	\label{gradient}
\end{equation}
where $m$ is the magnetization defined in (\ref{m}). This is a system of $N$ coupled non linear equations of degree $3^N$, thus, if we aspect at most $3^N$ solutions. Since the equations of the system (\ref{gradient}) are all equal, let us omit the index $i$, and consider the equation
\begin{equation}
    \phi^3-\phi-Jm=0.
		\label{phi3}
\end{equation}

\begin{figure}
\begin{center}
\includegraphics[width=0.235\textwidth]{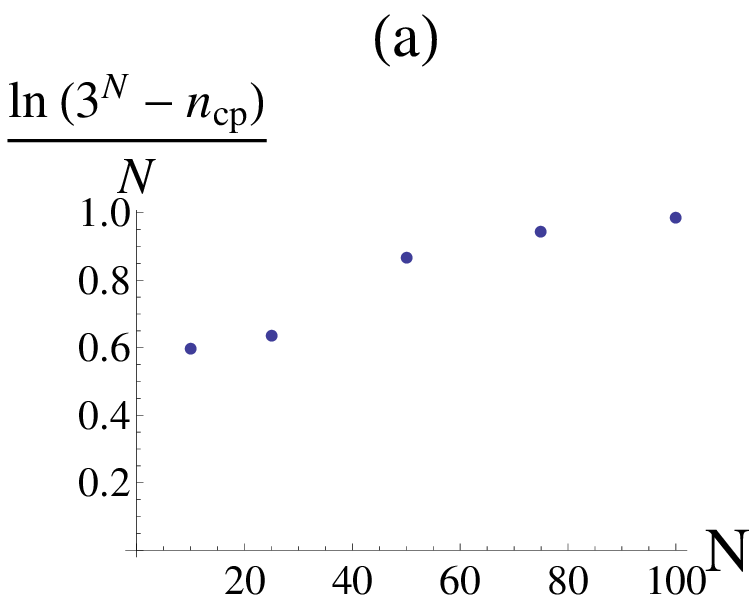}
\includegraphics[width=0.235\textwidth]{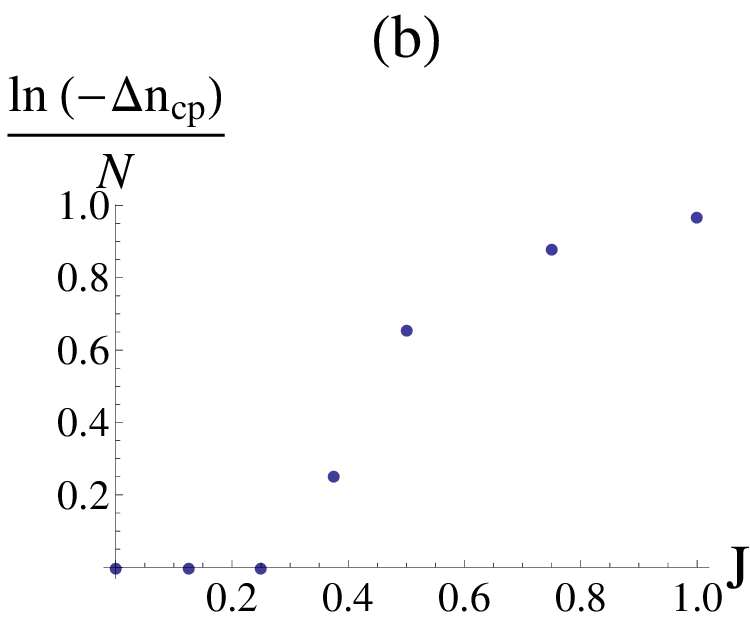}
\caption{Mean-field $\phi^4$ model (\ref{V}). (a) Estimate of the difference between $3^N$ and the total amount of critical points $n_{cp}$ as a function of $N$ for $J=0.5$. (b) $\Delta n_{cp}\equiv n_{cp}(J_i)-n_{cp}(J_{i-1})$, where the $J_i$'s are the plotted points, for $N=50$.  The total amount of critical points diminishes while increasing $J$ for $J>0.25$, up to $J\approx 0.25$ it equals $3^N$.}
\label{ncp}
\end{center}
\end{figure}

Consider the following cases. 

\smallskip
(i); $|Jm|>\frac{2}{3\sqrt{3}}$. The equation (\ref{phi3}) has one real solution.

\smallskip
(ii); $|Jm|\leq\frac{2}{3\sqrt{3}}$. The equation (\ref{phi3}) has three real solutions, two of them are coinciding in the limiting case '$=$'.

\smallskip
Case (i) is easier to treat because the system (\ref{gradient}) has an unique solution with components $\phi_i=\phi_0$, $i=1,\cdots,N$, where $\phi_0$ is solution of 
\begin{equation}
    \phi^3-(1+J)\phi=0,
\end{equation}
therefore, the solutions, with the respective potential values, are 
\begin{eqnarray}
   \phi_1=0,\quad v\left(\phi_1\right)=0,
		\\
	\phi_{2,3}=\pm\sqrt{1+J},\quad v\left(\phi_{2,3}\right)=-\frac{1}{4}(1+J)^2.
\end{eqnarray}

Case (ii). The solutions of the system (\ref{gradient}) are given by
\begin{equation}
    \phi^s=(\underbrace{\phi_1,\cdots,\phi_1}_{n_1}, \underbrace{\phi_2,\cdots,\phi_2}_{n_2},\underbrace{\phi_3,\cdots,\phi_3}_{N-n_1-n_2})
		\label{gensol}
\end{equation}
with all the permutations of $\phi_1$, $\phi_2$, and $\phi_3$, whose number is the multinomial coefficient
\begin{equation}
(N,n_1,n_2,N-n_1-n_2)!=\frac{N!}{n_1!n_2!(N-n_1-n_2)!}
\label{multi}
\end{equation}
for every choice of $n_1, n_2$ such that 
\begin{eqnarray}
0\leq n_1\leq N,
\\
0\leq n_2\leq N-n_1.
\label{n1n2}
\end{eqnarray}
Furthermore, $\phi_1$, $\phi_2$, and $\phi_3$ have to satisfy the constraint
\begin{equation}
    Nm=n_1\phi_i+n_2\phi_2+(N-n_1-n_2)\phi_3.
		\label{char}
\end{equation}
There are $\sum^{N+1}_{1}i=\frac{1}{2}(N+1)(N+2)$ independent equations of the form (\ref{char}). 

To summarize, for a given choice of $n_1, n_2$ we obtain, if there exist, some solutions of $m$ which yield solutions of the form (\ref{gensol}) with multiplicity $(N,n_1,n_2,N-n_1-n_2)!$ and with critical level $v(\phi^s(m))$. In the following sections we will see how to calculate the index of every critical point. We have limited to show the results up to $J=1$ because some numerical problems makes the results for $J>1$ not entirely reliable. 

\begin{figure}
\begin{center}
\includegraphics[width=0.235\textwidth]{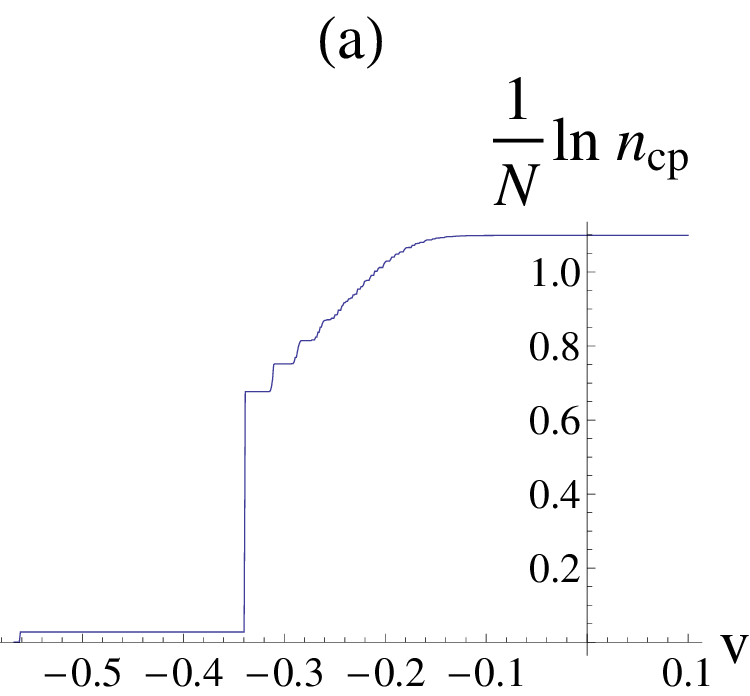}
\includegraphics[width=0.235\textwidth]{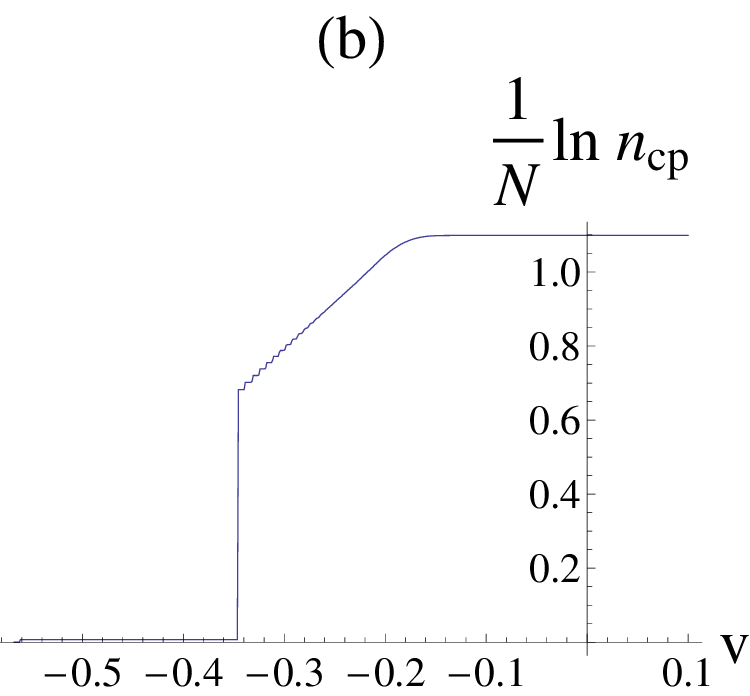}
\includegraphics[width=0.235\textwidth]{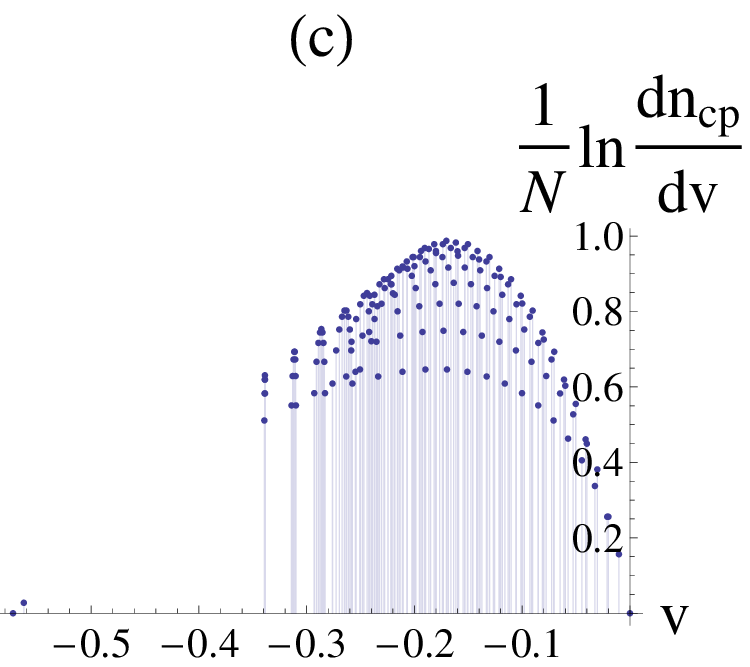}
\includegraphics[width=0.235\textwidth]{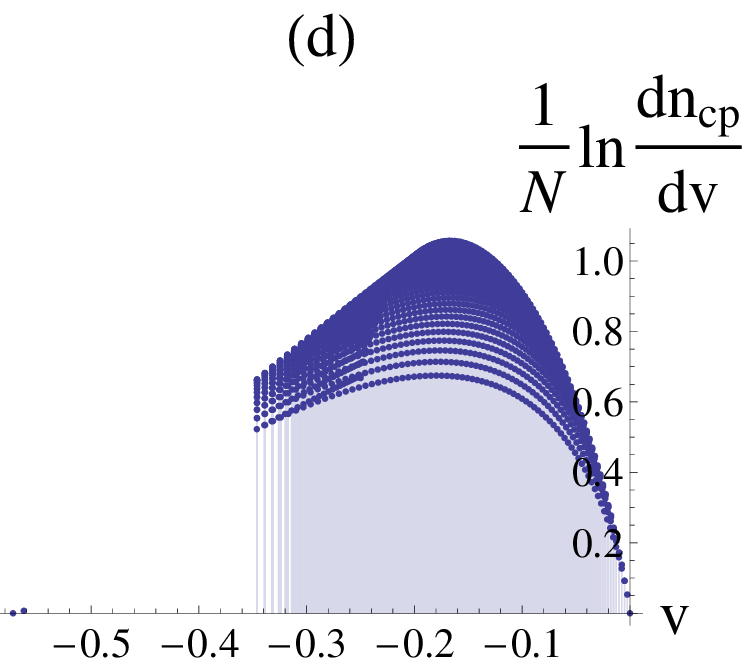}
\caption{Mean-field $\phi^4$ model (\ref{V}) with $J=0.5$. (a), (c) Density of the logarithmic amount of critical points $n_{cp}$ and logarithmic density of the amount of critical points, respectively, as functions of the potential density $v$ for $N=25$. (b), (d) The same as in (a), (c) for $N=100$.}
\label{phi4_ncp}
\end{center}
\end{figure}

In Fig. \ref{phi4_ncp} the amount of critical points and their density are reported as functions of the potential $v$ for two values of $N$. That quantities are scaled by the factor $1/N$, so that the graphs show a very similar shape while varying $N$. This can be explained by the fact that the amount of critical points goes exponentially with $N$ at fixed values of $J$.

It is easy to prove analytically that all the critical levels of the potential are bounded from above by zero. Start by observing that $V$ can be written as
\begin{equation}
   V=\sum^{N}_{i=1}\left(\phi_i\left(\phi^3_i-\phi_i-Jm\right)-\frac{\phi_i^4}{4}\right),     
\end{equation}
if $\phi^s$ is a solution of $\nabla V=0$, the conclusion is immediately reached. In \cite{km} it has been shown the same result for the $2D$ $\phi^4$ model with nearest-neighbors interaction. 

\subsubsection{Index of the critical points}

\begin{figure}
\begin{center}
\includegraphics[width=0.235\textwidth]{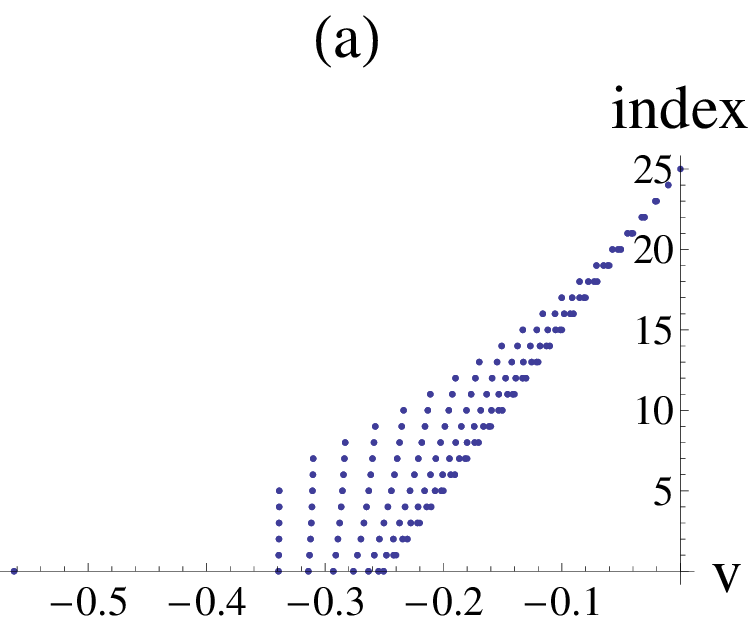}
\includegraphics[width=0.235\textwidth]{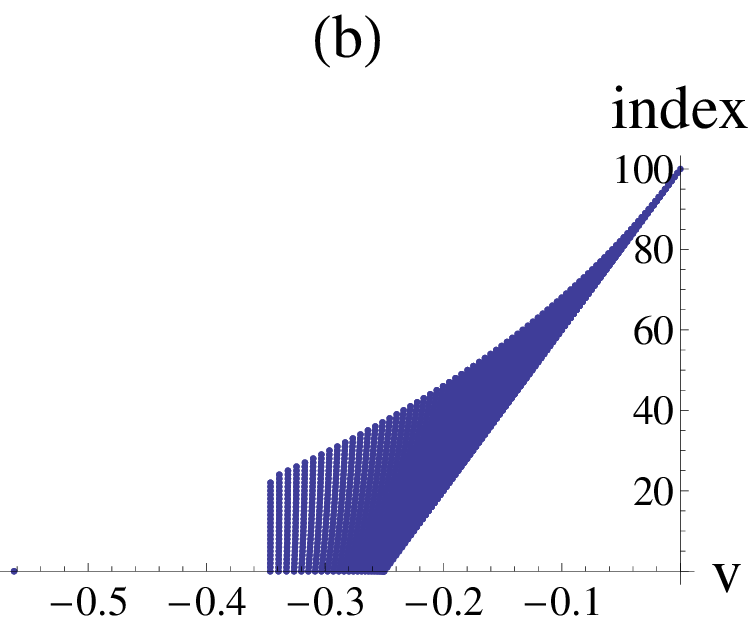}
\includegraphics[width=0.235\textwidth]{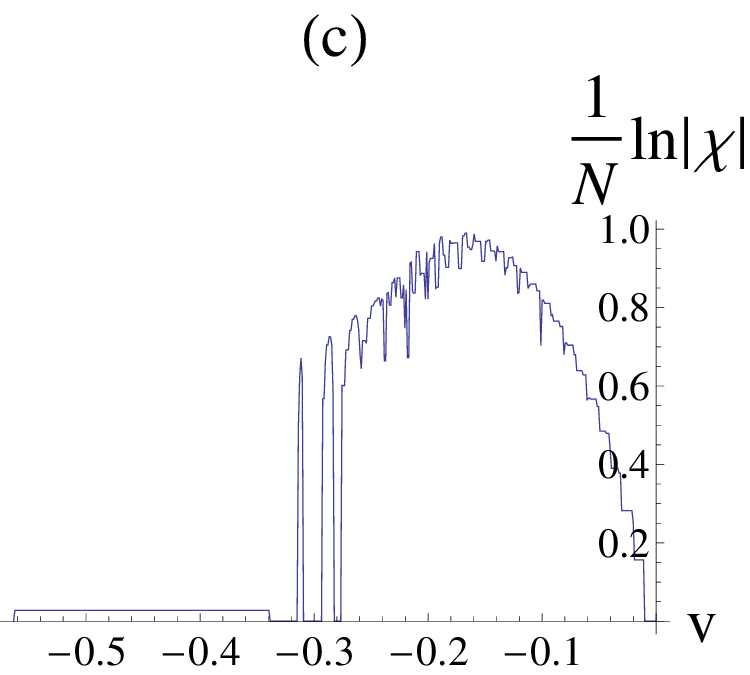}
\includegraphics[width=0.235\textwidth]{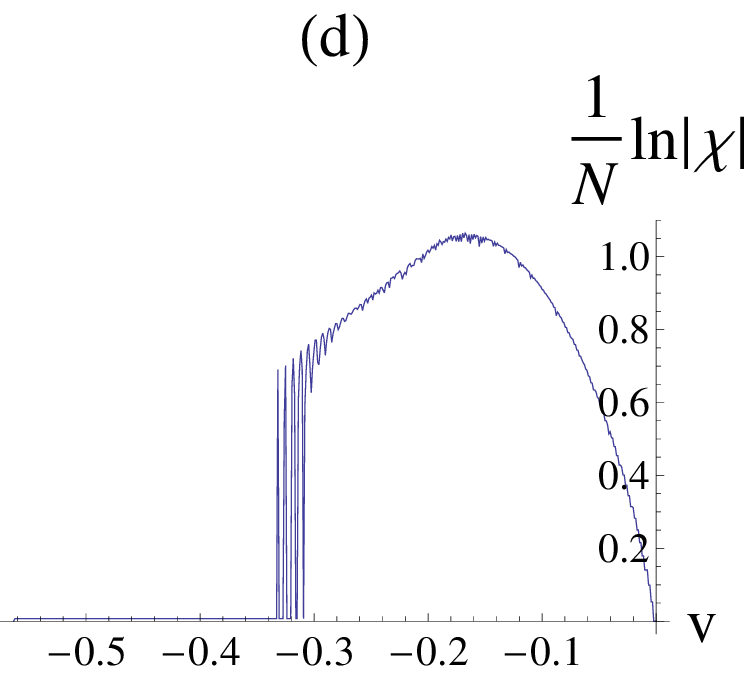}
\caption{Mean-field $\phi^4$ model (\ref{V}) with $J=0.5$. (a), (c) Index of the critical points and specific logarithmic modulus of the Euler characteristic, respectively, as functions of the potential density $v$ for $N=25$. (b), (d) the same as (a), (c) for $N=100$.}
\label{phi4_indexchi}
\end{center}
\end{figure}

\begin{figure}
\begin{center}
\includegraphics[width=0.235\textwidth]{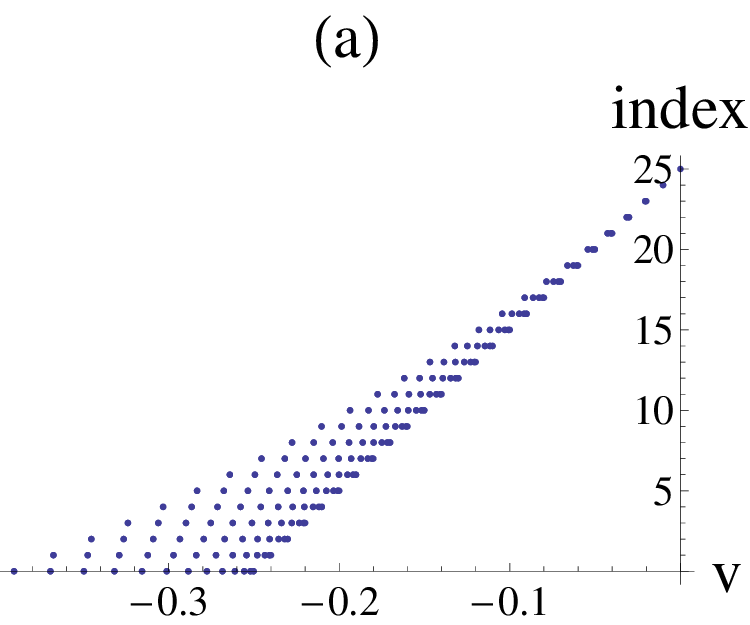}
\includegraphics[width=0.235\textwidth]{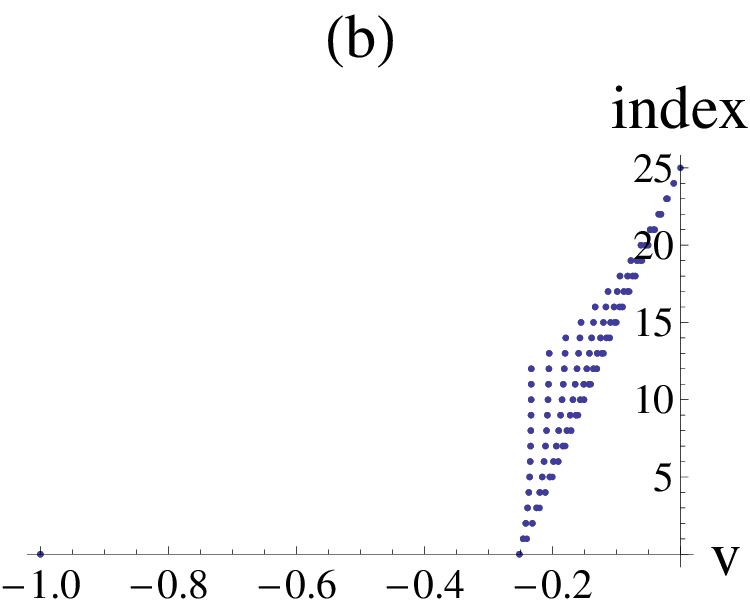}
\includegraphics[width=0.235\textwidth]{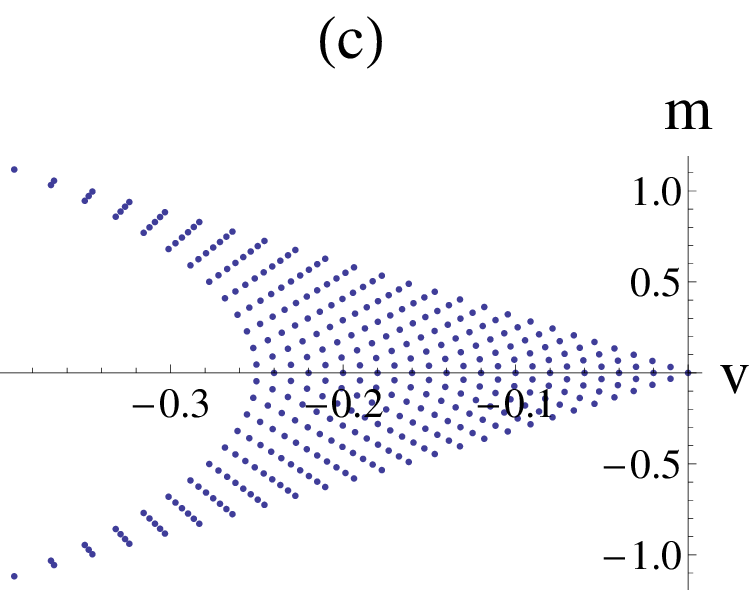}
\includegraphics[width=0.235\textwidth]{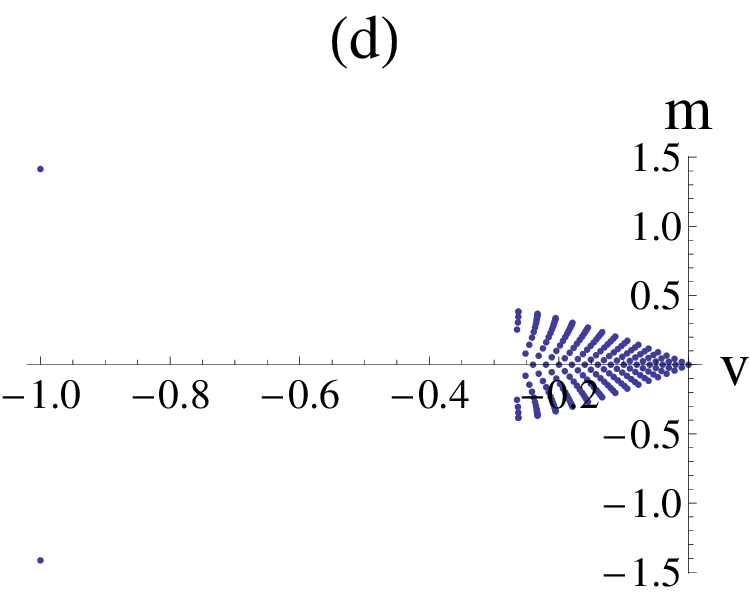}
\caption{Mean-field $\phi^4$ model (\ref{V}) with $N=25$. (a), (c) Index and magnetization, respectively, of the critical points as functions of the potential density $v$ for $J=0.25$. (b), (d) The same as in (a), (c) for $J=1$.}
\label{phi4_index}
\end{center}
\end{figure}

In Morse theory the index of a critical point is the number of negative eigenvalues of the Hessian matrix $H$, which for the potential (\ref{V}) takes the form
 \begin{equation}
   H_{ij}=\frac{\partial^2 V}{\partial\phi_i\partial\phi_j}=\left(3\phi_i^2-1\right)\delta_{ij}-\frac{J}{N}.         
\end{equation}
$H$ can be written as $H=D+B$, with
\begin{eqnarray}
D_{ij}=(3\phi_i^2-1)\delta_{ij},
		\\
B=-\frac{J}{N}U,
\end{eqnarray}
where $U$ is the matrix whose elements equal $1$. The eigenvalues of $D$ read directly on its diagonal, while $B$ has an unique non-zero eigenvalue of value $N$ because its rank is $1$. Because of the form of $H$, we can apply a results based on the Wilkinson theorem which guarantees that, to get the number of negative eigenvalues of $H$, if we take the number of the negative eigenvalues of $D$ we make at most an error of $\pm 1$. For more details we refer to \cite{ccp1,ccp2}. Some results are shown in Fig. \ref{phi4_indexchi} and \ref{phi4_index}.

\subsubsection{Euler characteristic}

The Euler characteristic $\chi$ is a topological invariant, i.e. a function of a manifold which does not change value if the manifold is deformed without varying its topology. $\chi$ is defined by the Betti numbers $b_k$ \cite{pettini}, Morse theory allows us to calculate it for the $M_{v,N}$'s by the relation
\begin{equation}
   \chi(v,N)\equiv\sum^{N}_{k=0}(-1)^k b_k(M_{v,N})=\sum^{N}_{k=0}(-1)^k \mu_k(M_{v,N}),         
\end{equation}
where the Morse number $\mu_k$ is the number of critical points of $M_{v,N}$ that have index $k$. In Fig. \ref{phi4_indexchi} we have plotted $1/N\ln |\chi(v)|$ because it approximately does not depend on $N$.  This is due to the fact that the number of critical points grows exponentially with $N$. The modulus appears because $\chi$ is in general an oscillatory function of $v$ above and below zero. At $v>0$ we have found $\chi(v)=1$. This is coherent with the fact all the critical levels are below zero, because, as a consequence, $M_{v,N}$ for $v>0$ is homeomorphic to an $N$-ball which has $\chi=1$. 

\section{$\phi^4$ model without interaction}
\label{phi4noint}

\subsection{Canonical thermodynamic}

\begin{figure}
\begin{center}
\includegraphics[width=0.235\textwidth]{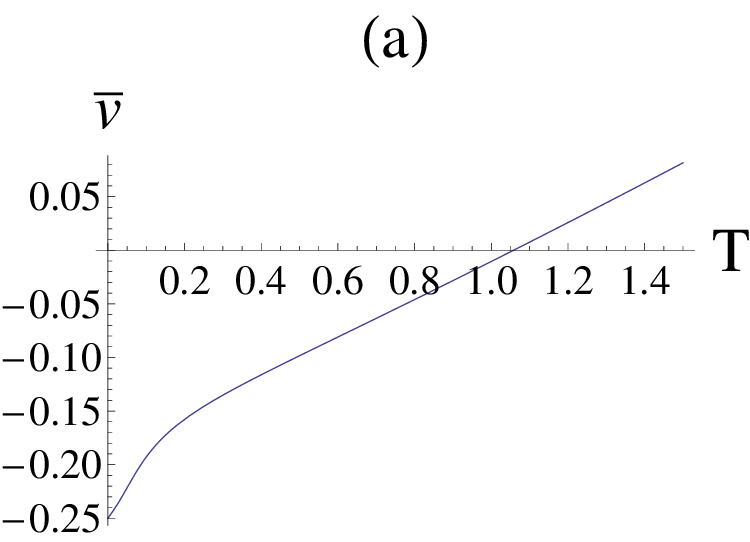}
\includegraphics[width=0.235\textwidth]{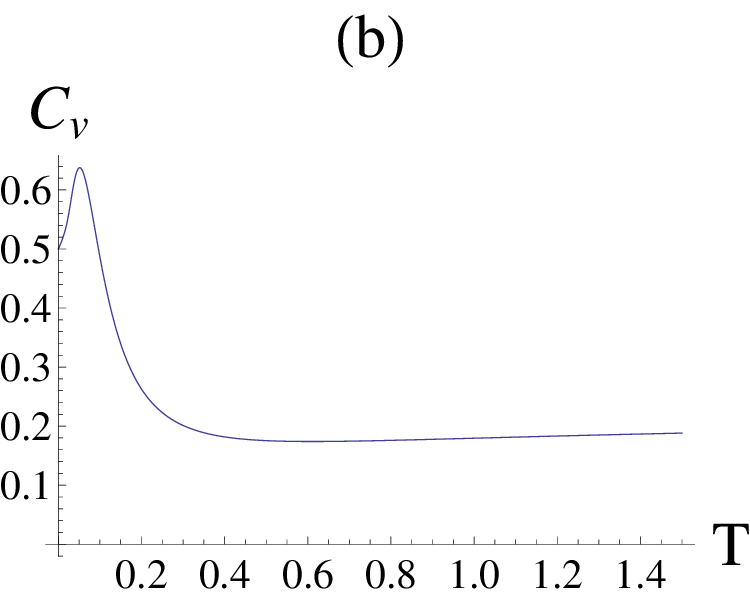}
\caption{$\phi^4$ model without interaction (\ref{Vloc}). (a) Average potential density $\overline{v}$ as a function of the temperature $T$. (b) Specific heat $C_v$ as a function of $T$.}
\label{phi4noint_vCvT}
\end{center}
\end{figure}

In order to make a confront with a model without SBPT, we have studied the $\phi^4$ model without interaction. The potential is nothing but that of the Hamiltonian (\ref{V}) when the interacting terms have been deleted, i.e. as $J=0$ is set. The configurational partition function is 
\begin{equation}
   Z_c=\int\prod^{N}_{i=1}d\phi_i\, e^{-\beta\sum^{N}_{i=1}V(\phi_i)} =\left(\int d\phi\,e^{-\beta V(\phi)}\right)^N,
\end{equation}
where $V(\phi)$ is the local potential (\ref{vlocal}). The analytic solution can be obtained by the following integral
\begin{equation}
   \int^{\infty}_{0}dz\,z^{\nu-1}e^{-\gamma z-\alpha z^2}=(2\alpha)^{-\frac{\nu}{2}}\Gamma (\nu)e^{\frac{\gamma^2}{8\alpha}}D_{-\nu}\left(\frac{\gamma}{\sqrt{2\alpha}}\right),
\end{equation}
where $D_{\nu}(z)$ are parabolic cylinder functions. After some algebraic manipulation, we get 
\begin{equation}
   Z_c=\left(\left(\frac{\beta}{2}\right)^{-\frac{1}{4}}\Gamma \left(\frac{1}{2}\right)e^{\frac{\beta}{8}}D_{-\frac{1}{2}}\left(-\sqrt{\frac{\beta}{2}}\right)\right)^N.
\end{equation}
No SBPT can occur because the thermodynamic function do not depend on $N$, so that the thermodynamic limit cannot generate any emergent behaviour. The average potential density $\overline{v}$ and specific heat $C_v$ as functions of the temperature $T$ are reported in Fig. \ref{phi4noint_vCvT}

\subsection{Topology of the submanifolds $M_{v,N}$'s}
\label{phi4nointtop}

The potential of the Hamiltonian (\ref{V}) can be written as $V=V_{loc}+V_{int}$, so that the potential of our model is
 \begin{equation}
   V_{loc}=\sum^{N}_{i=1}\left(-\frac{\phi_i^2}{2}+\frac{\phi_i^4}{4}\right).
	\label{Vloc}
\end{equation}
As already made for the m-f case, we have to solve $\nabla V_{loc}=0$, which is the following system
\begin{equation}
   \phi^3_i-\phi_i=0\quad\quad i=1,\cdots,N.
	\label{gradientnoint}
\end{equation}
It is immediate to see that the solutions are of the form
\begin{equation}
    \phi^s=(\underbrace{\phi_1,\cdots,\phi_1}_{n_1}, \underbrace{\phi_2,\cdots,\phi_2}_{n_2},\underbrace{\phi_3,\cdots,\phi_3}_{N-n_1-n_2}),
		\label{gensolnoint}
\end{equation}
where, without loss of generality, $\phi_1=1$, $\phi_2=-1$ and $\phi_3=0$. $n_1$ and $n_2$ follow the same rule (\ref{n1n2}). The multiplicity of the solutions for given $n_1, n_2$ is the same of the m-f $\phi^4$ model (\ref{multi}). 

\begin{figure}
\begin{center}
\includegraphics[width=0.235\textwidth]{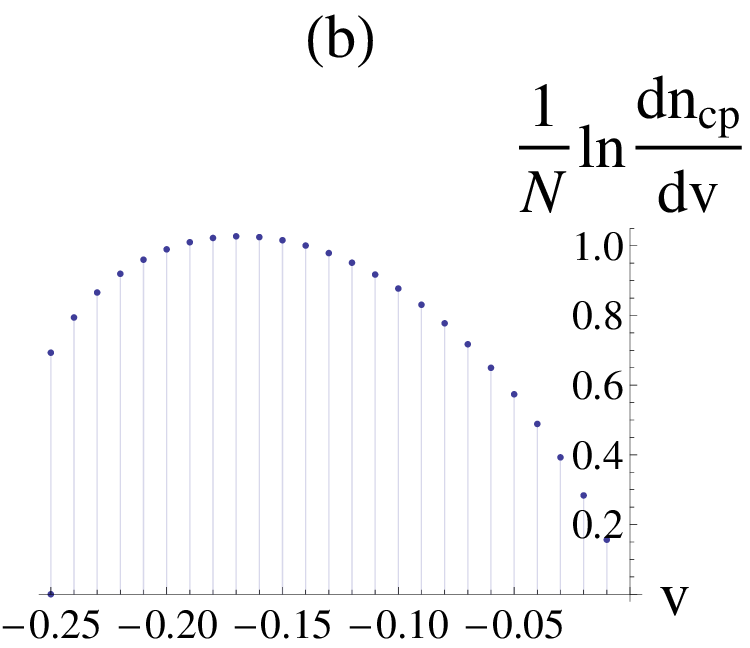}
\includegraphics[width=0.235\textwidth]{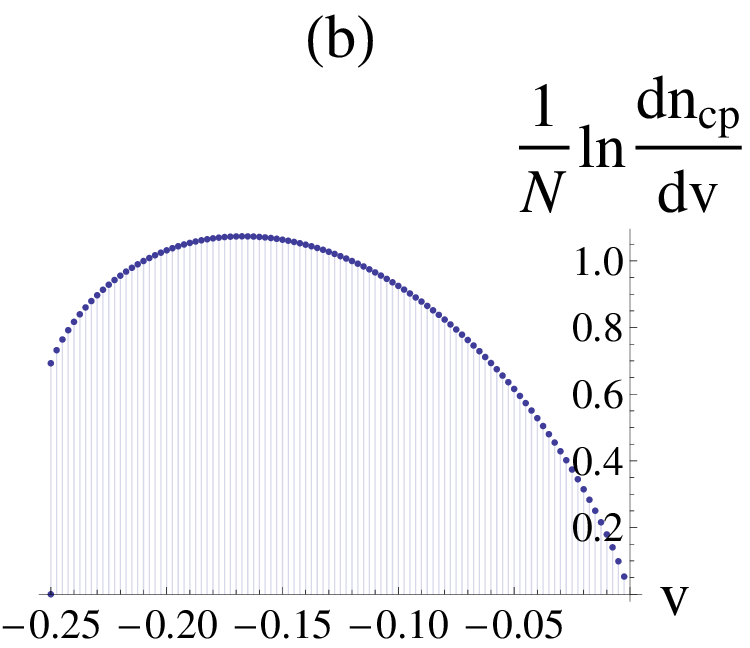}
\includegraphics[width=0.235\textwidth]{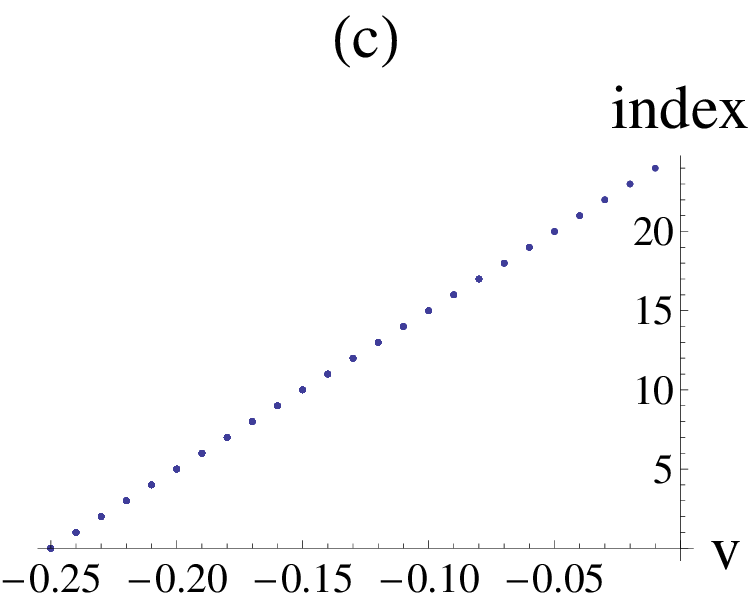}
\includegraphics[width=0.235\textwidth]{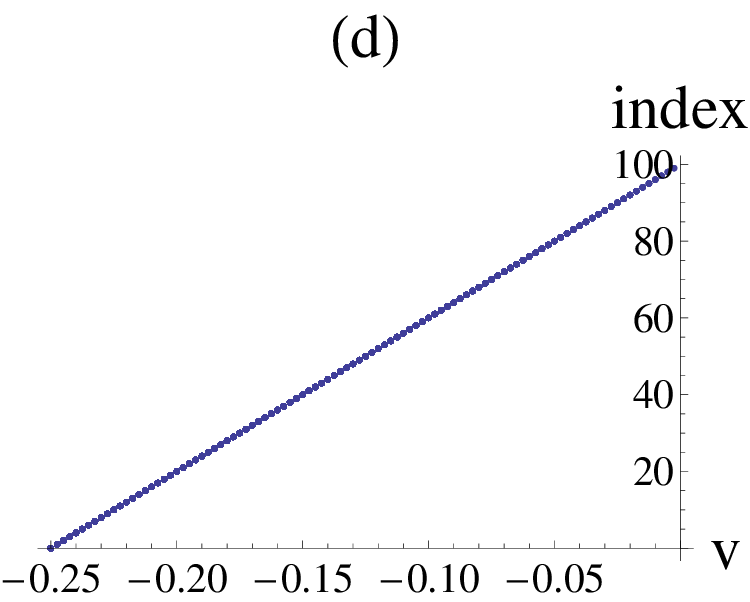}
\caption{$\phi^4$ model without interaction (\ref{Vloc}). (a) Logarithmic density of critical points per degree of freedom at $N=25$ as a function of the potential density $v$. (c) Indexes of the critical points at $N=25$ as a functions of $v$. (b), (d) The same as in (a), (b) at $N=100$. Increasing $N$ does not entail any qualitative difference. The total amount of critical points is fixed at $3^N$.}
\label{phi4noint_dlicpv}
\end{center}
\end{figure}

\begin{figure}
\begin{center}
\includegraphics[width=0.235\textwidth]{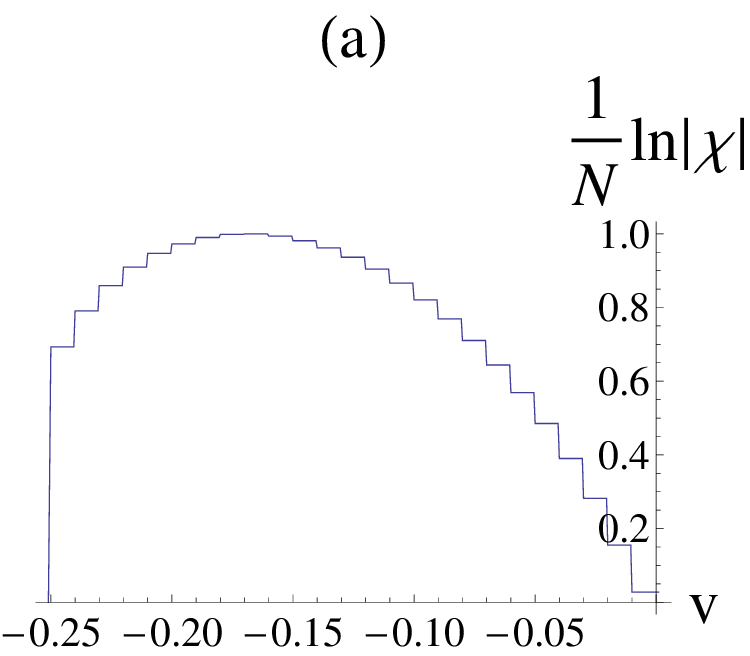}
\includegraphics[width=0.235\textwidth]{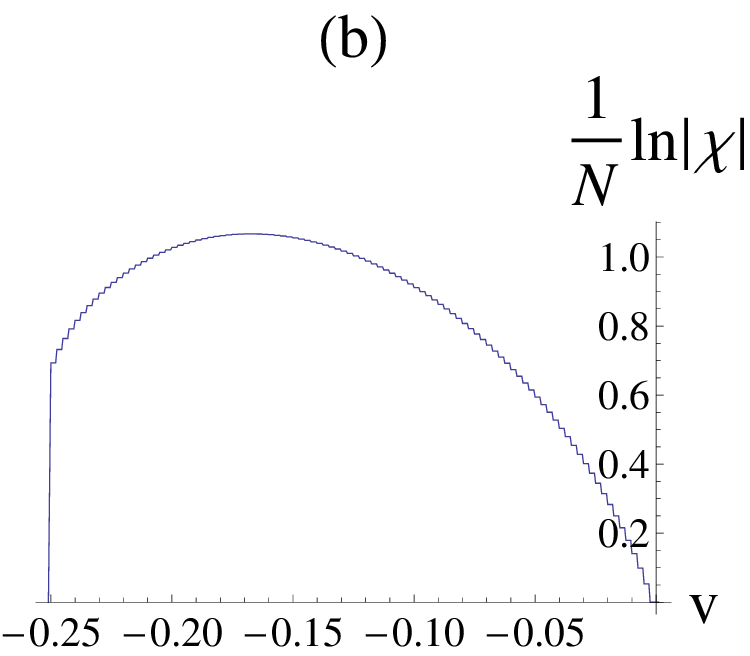}
\caption{$\phi^4$ model without interaction (\ref{Vloc}). (a) Logarithmic absolute value of the Euler characteristic $\chi$ per degree of freedom at $N=25$ as a function of the potential density $v$. (b) The same as in (a) at $N=100$.}
\label{phi4noint_echarv}
\end{center}
\end{figure}

The Hessian matrix takes the form
 \begin{equation}
   H_{ij}=\frac{\partial^2 V}{\partial\phi_i\partial\phi_j}=(3\phi_i^2-1)\delta_{ij},     
\end{equation}
so that the index of the critical points is simply $N-n_1-n_2$. In Fig. \ref{phi4noint_dlicpv} the logarithmic density of critical points per degree of freedom and their indexes are reported  as functions of the potential for two values of $N$.
For the Euler characteristic, all proceeds as for the m-f $\phi^4$ model (see Fig. \ref{phi4noint_echarv}). 

\section{Discussion of the results}
\label{results}

For convenience, in what follows we will consider the $\Sigma_{v,N}$'s instead of the $M_{v,N}$'s, but they are perfectly equivalent for our purpose.
From a topological viewpoint, it is convenient to divide the range of the accessible $v$'s of the m-f $\phi^4$ model in three regions: 

\medskip
(i)  $[v_{min},v')$; the $\Sigma_{v,N}$'s are equivalent to the disjointed union of two $N$-spheres. $v_{min}=-(1/4)(1+J)^2$, while $v'$ is a functions of $N$ of which some points are plotted in Fig. \ref{v'}.

\smallskip
(ii) $[v',0]$; the topology of the $\Sigma_{v,N}$'s are dramatically intricate. Simplifying the situation, the interval $[v',0]$ plays the role of a critical level which divides region (i) from region (iii).

\smallskip
(iii) $(0,+\infty)$; the $\Sigma_{v,N}$'s are equivalent to an $N$-sphere.

\medskip
$v'$ is bounded from above by $-1/4$. This is shown in Fig. \ref{v'} at least up to $J=1.1$, furthermore it can be analytically proven in the following way. 
Among the solutions (\ref{gensolnoint}) of $\nabla V_{loc}=0$, where $V_{loc}$ is defined in (\ref{Vloc}), consider the $\phi^s_{0}$'s for $n_1=0,\cdots,N$, and $n_2=n_1$. The corresponding magnetization is vanishing because it is given by $m=(n_1-n_2)/N$. Since
\begin{equation}
\nabla V=\nabla V_{loc}-JNm\nabla m,
\end{equation}
we see that the $\phi_{0}^s$'s are solutions even for $\nabla V=0$. The corresponding value of the potential is $v=-1/(4N)(n_1+n_2)$. Now distinguish two cases: $N$ even and $N$ odd. If $N$ is even, for $n_1=N/2$, $v=-1/4$, so that the latter is an upper bound of $v'$. If $N$ is odd, for $n_1=\left[N/2\right]$, where $[\cdot]$ is the integer part, $v=-1/4+1/(4N)$, so that $-1/4$ is an upper bound of $v'$ even this case. 

We conclude that, neither the growing of $N$, nor the growing of $J$ can restrict the critical region (ii). In a further paper we will see how it is possible to reduce it to an unique critical level with an unique critical point of index $1$, i.e. a saddle.

Since as $J$ increases the total amount of critical points tends to reduce from a maximum of $3^N$ for $J<0.25$ (see Fig. \ref{ncp}, \ref{phi4_index}), we are led to conjecture that there exists $J_0$, eventually dependent on $N$, such that for $J>J_0$ the critical points are the only ones with vanishing magnetization, i.e. with $n_2=n_1$. In other words, the degeneracy on $n_2$ would be removed by values of $J$ large enough. Our conjecture is reinforced by a similar result that has been found out in \cite{km,mhk} for the $2D$ $\phi^4$ model with nearest-neighbors interaction via numerical analysis.

In what follows we will analyze these results at the light of two theorems recently proven.

\subsection{Theorems on a sufficiency topological condition for $\mathbb{Z}_2$-SBPT}

In \cite{bc} two straightforward theorems on a sufficiency geometric-topological condition for $\mathbb{Z}_2$-SBPTs have been shown. For the sake of clarity, in the following considerations we will simplify a little bit the picture. 

Consider an $N$ degrees of freedom Hamiltonian system with a $\mathbb{Z}_2$-symmetric potential bonded from below. The sufficiency condition contained in Theorem 1 is as follows: if the $\Sigma_{v,N}$'s are made by two disjointed connected components $A_+$, $A_-$ for $v\in [v_{min},v')$ such that $A_+$ is the image of $A_-$ under the $\mathbb{Z}_2$ symmetry for any $N$, then the symmetry is broken for $\overline{v}\in [v_{min},v')$, where  $\overline{v}(T)$ is the average potential (density) selected by the temperature $T$. Furthermore, if there exists a critical potential $\overline{v}_c$ at which the $\mathbb{Z}_2$ symmetry breaks, then $\overline{v}_c\geq v'$ has to hold.

Again, Theorem 2 states that, if the $\Sigma_{v,N}$'s are topologically equivalent to an $N$-sphere and ergodic for $v>v''$ with $v''>v'$, then the $\mathbb{Z}_2$ symmetry is intact for $\overline{v}>v''$, so that a phase transition, meant as a non-analytic point in the magnetization, has to occur as a consequence. 

\begin{figure}
\begin{center}
\includegraphics[width=0.235\textwidth]{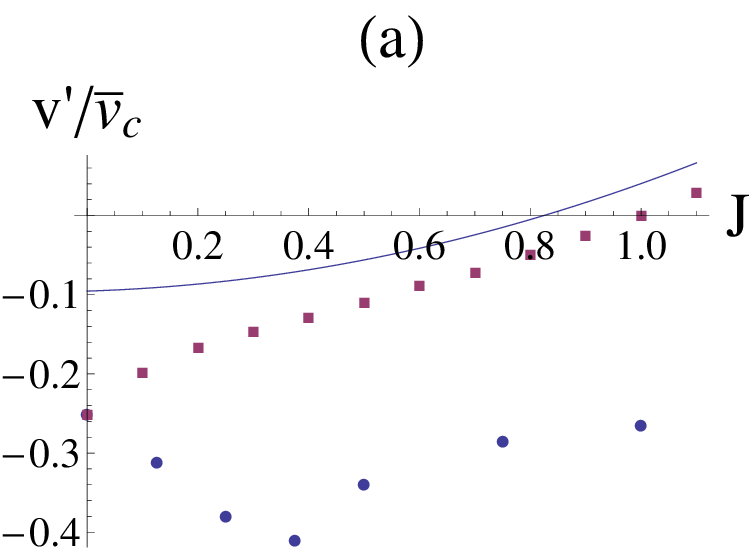}
\includegraphics[width=0.235\textwidth]{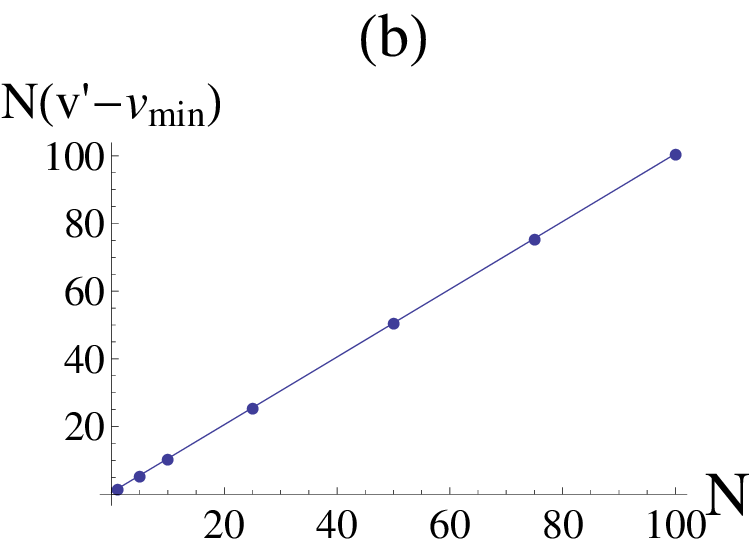}
\caption{(a) Critical average potential $\overline{v}_c$ (squares) and the nearest critical level $v'$ (disks) above the global minimum as functions of the coupling constant $J$. The line is the asymptotic expression (\ref{kastner}) of $\overline{v}_c(J)$ at large $J$. (b) The interval of $V$-values starting from the absolute minimum at which the $M_{v,N}$ are made by two disjointed $N$-spheres as a function of the number of degrees of freedom $N$. The line is a linear fit.}
\label{v'}
\end{center}
\end{figure}

Now, will see that the m-f $\phi^4$ model satisfies the hypotheses of Theorem 1, at least for the values of $N$ considered here, anyway, we have no reasons to doubt that this can hold for every $N$. Let us see why. If a double-well potential has the gap between the wells proportional to $N$, then it satisfies the hypotheses of Theorem 1. Indeed, for the values of the potential comprised between the absolute minimum and the gap between the wells the $\Sigma_{v,N}$'s are topological equivalent at least to two disjointed connected components which are non-symmetric under $\mathbb{Z}_2$ singularly considered. Fig. \ref{v'} shows that the difference $v'-v_{min}$ is proportional to $N$, at least for the $N$'s considered here. The gap is bounded from above by $0$, so that it is proportional to $N$ because $v_{min}$ goes as $-N$. We can so conclude that the m-f $\phi^4$ model satisfied the hypotheses of Theorem 1.

Fig. \ref{v'} shows that the thermodynamic critical potential $\overline{v}_c\geq v'$, as requested by Theorem 1, at least for $0<J\leq 1$. To demonstrate that this also holds for $J>1$, we resort to the result obtained in \cite{hk} which provides that 
\begin{equation}
\overline{v}_c=a^2J^2-\left(2a^2-\frac{1}{4}\right)J+\left(\frac{5a^2}{4}-\frac{3}{8}+\frac{1}{64a^2}\right)+O\left(\frac{1}{J}\right), 
\label{kastner}
\end{equation}
where $a=\Gamma\left(3/4\right)/\Gamma\left(1/4\right)$.

Let us try to emphasize the very difference between the model with $\mathbb{Z}_2$-SBPT and the one without.
The potential of the latter has $2^N$ global minima. The effect of the interaction is that the two minima, whose coordinates in configuration space are $\pm (1,\cdots,1)$, sink with respect to all the others becoming global and transforming the others into local minima. Furthermore, the distance between $v_{min}$ and the nearest critical level $v'$ is proportional to $N$. We have no analytical proof of this, but the results support this conjecture. This is the true difference between the model with interaction and the one without. In other words, the potential becomes double-well-shaped with a gap proportional to N thanks to the interaction. For the sake of precision, it is not properly a double-well because otherwise it had only three critical points: two global minima and a saddle, but the large amount of local minima does not affect the generating-mechanism of the $\mathbb{Z}_2$-SBPT.

Now we will discuss about Theorem 2. If the ergodicity held for all the $\Sigma_{v,N}$'s belonging to the region (iii), then $\overline{v}_c$ would be located in the region (ii) for any $J$. But since this is not true, the ergodicity cannot be resort to explain the $\mathbb{Z}_2$-SB for $v\in (0,\overline{v}_c)$, despite the fact that the $\Sigma_{v,N}$'s are topologically equivalent to an $N$-sphere. This means that, besides to the topological one, another $\mathbb{Z}_2$-SB-mechanism has to be at work. 

About this issue, in \cite{b3} such a mechanism for $\mathbb{Z}_2$-symmetric systems has been put forward starting from a particular shape of the $\Sigma_{v,N}$'s which can be intuitively defined dumbbell-shaped. We remark that this $\mathbb{Z}_2$-SB-mechanism has nothing to do with the topology of the $\Sigma_{v,N}$'s. Here we hypothesize that the $\Sigma_{v,N}$'s under the critical thermodynamic potential $\overline{v}_c$ are dumbbell-shaped, so that the $\mathbb{Z}_2$ symmetry is broken, and above $\overline{v}_c$ instead they are not, so that the $\mathbb{Z}_2$ symmetry is intact. 
We note that the topological mechanism proposed in \cite{bc} is nothing but a particular case of the more general one based on dumbbell-shaped $\Sigma_{v,N}$'s.

\subsection{Euler characteristic}

In \cite{ccp,ccp1,ccp2} the $XY$ model in m-f version and without interaction has been studied following the same procedure of this paper. We recall that the m-f $XY$ model undergoes a second-order $O(2)$-SBPT with classical critical exponent, as the m-f interaction requires. The thermodynamic critical average potential is in correspondence of a topological critical level at which a topological change occurs with critical points of all possible indexes, and the graph of the modulus of the Euler characteristic $1/N\ln|\chi|$ shows a jump. Obviously, $1/N\ln|\chi|$ has a jump at any critical level, but provided that $N$ is large enough it can be approximated by a continuous function. At converse, in the model without interaction $1/N\ln|\chi|$ has a continuous shape everywhere.

These results have suggested a strong relation between SBPTs and topological changes, but this scenario is not confirmed by the results of the m-f $\phi^4$ model founded out here. $1/N\ln|\chi|$ shows a jump at $v'$ followed by a very intricate shape and by an angular point at $v=0$. $v'$ never corresponds to the thermodynamic critical average potential, while $v=0$ corresponds only for the particular value $J=1$. Worst, this shape is showed even by the model without interaction where no SBPT occurs, even though the graph is more regular. We conclude that the scenario depicted by the m-f $XY$ model is a particular case without any particular significance from a viewpoint of the relation between SBPT and topology of configuration space.

\section{Concluding remarks}

In this paper we have analytically characterized the topology of the $M_{v,N}$'s of the m-f $\phi^4$ model and the same model without interaction by means of Morse theory. Then, we have tried to discover any possible link with the $\mathbb{Z}_2$-SBPT occurring in the m-f $\phi^4$ model. The critical levels remain confined in an interval of the potential which is $[-1/4,0]$ for the model without interaction and $[v',0]$ for the m-f model, where $v'\rightarrow -(1/4)^-$ as $J$ increases. 

Despite the huge amount of critical points growing as $3^N$ for $J$ small enough, but less and less at the increasing of $J$, in our opinion all that critical points do not have any particular significance except that to separate the region (i), where the $M_{v,N}$'s are homeomorphic to two disjointed $N$-balls, from the region (iii), where $M_{v,N}$'s are homeomorphic to an $N$-ball alone. This is true only in the m-f $\phi^4$ model, where the potential has turned out to be a double-well with the two global minima separated by a gap proportional to $N$. This is the effect of the m-f interaction and it is the very peculiar difference between the two models. Indeed, the global minima of the model without interaction are $2^N$ and remain located at $v=-1/4$. 

In our opinion, this is the main result of this study. The sufficiency condition for $\mathbb{Z}_2$-SBPT stated in \cite{bc} has already been found at work in various toy models \cite{b4,b,b1}, but here for the first time on a physical model with continuous variables. We wonder how this scenario may be generalized to the short-range case. A double-well potential with gap proportional to $N$ implies a non-concave graph of the microcanonical entropy $s(v,m)$, while in a short-range system it must be concave or non-strictly concave in the presence of a PT. We conjecture that in a short-range system the gap increases slowly than $N$. Indeed, it can be shown that for a short-range $d$-dimensional Ising model the gap is proportional to $N^{(d-1)/d}$ in the limit of large $N$. 

Nevertheless, in \cite{mhk} the authors have found the potential of the nearest-neighbor $2D$ $\phi^4$ model to have only three critical points, i.e. a saddle at the $0$-level set and two global minima at a $v$-level set proportional to $-N$, for values of the coupling constant $J$ large enough. Let $J_0(N)$ be the minimum value of $J$ for which the critical points are only three at fixed $N$. It turns out that $J_0(N)$ allegedly goes as $N^2$, so that, starting from $J_0(N)$ and $N$, new critical points, whose critical levels belong to $(v_{min},0)$, necessarily arise while increasing $N$. This is not a proof that the gap cannot be proportional to $N$, but if this were not true, then the gap should be necessarily proportional to $N$ even in the nearest-neighbor $2D$ $\phi^4$ model.

Another remarkable fact is that the thermodynamic critical potential $\overline{v}_c$ can also be placed in correspondence of $\Sigma_{v,N}$'s which are topological equivalent to an $N$-sphere. This means that a further $\mathbb{Z}_2$-SB mechanism must exist besides the topological one working for $v\in [v_{min},v')$. In other words, the ergodicity of a $\Sigma_{v,N}$ cannot be guaranteed by the only assumption that it is homeomorphic to an $N$-sphere, as is already arisen from the $2D$ $\phi^4$ model. We cannot even be sure that the topological mechanism acts in $v\in [v',0]$ because we cannot apply the sufficiency condition in \cite{bc}. A possible solution is put forward in \cite{b3}, where it is shown that the $\mathbb{Z}_2$-SB can be entailed by dumbbell-shaped $\Sigma_{v,N}$'s. In this scenario, the thermodynamic critical potential is exactly in correspondence with a particular $\Sigma_{v,N}$ acting as a divider between the dumbbell-shaped $\Sigma_{v,N}$'s and those that are not, for which the $\mathbb{Z}_2$ symmetry is intact.

To conclude, we hope that this work may give useful hints to deepen our understanding of the role of geometry and topology of configuration space in phase transitions phenomenon. A natural development may be to research sufficiency conditions for other symmetry groups, e.g. O(2), O(3),... and so on, based on the concepts here discussed.

\appendix

\section{A glance at Morse theory}
\label{Mt}

Morse theory links the topology of a given manifold $M$ with the properties of the critical points of smooth functions defined on it. Two manifold $M$ and $M'$ are topologically equivalent if they can be smoothly deformed one into the other, i.e. if there exists a \emph{diffeomorphism} $\psi$ that maps $M$ into $M'=\psi(M)$. Here we consider only compact, finite-dimensional manifolds, but most of the results can be extended to non-compact manifolds. The key ingredient of Morse theory is to consider the manifold $M$ as decomposed into the \emph{level sets} of a function $f$. We recall that the $v$-level set of $f: M\rightarrow\mathbb{R}$ is the set
\begin{equation}
f^{-1}(a)=\{x\in M: f(x)=a\}.
\end{equation}
M being compact, any function $f$ has a minimum $f_m$ and a maximum $f_M$, so that one can be build $M$ starting from $f^{-1}(f_m)$ and then adding continuously to it all other level surfaces up to $f^{-1}(f_M)$. Further, we define the 'part of M below $a$' as
\begin{equation}
M_a=\{x\in M: f(x)\leq a\}.
\end{equation}
As $a$ is varied between $f_m$ and $f_M$, $M_a$ describes the whole manifold $M$. 

For our purposes, we need to restrict the class of functions into the class of \emph{Morse functions}, which are defined as follows. A point $x_c$ is called a \emph{critical point} of $f$ if $\nabla f(x_0)=0$, while the value $f(x_0)$ is called a \emph{critical value}. $f$ is called a Morse function on $M$ if its critical points are non-degenerate, i.e., if the Hessian matrix of $f$ at $x_c$, whose elements in local coordinates are
\begin{equation}
H_{ij}=\frac{\partial^2 f}{\partial x_i\partial x_j},
\end{equation}
has rank $n$, where $n$ is the dimension of $M$. As a consequence, one can prove that the critical points $x_c$ of a Morse function, and also its critical values, are isolated. It can be proved also that the set of the Morse functions are dense in the space of the smooth functions from $M$ to $\mathbb{R}$. A level set is called \emph{critical level} if there is at least a critical point belonging to it. 

If the interval $[a,b]$ contains no critical values of $f$, then the topology of $f^{-1}([a,v])$ does not change for any $v\in (a,b]$. This result is sometimes called the \emph{noncritical neck theorem}. 

If the interval $[a,b]$ contains critical values, the topology of $f^{-1}[a,v]$ changes in correspondence with the critical values themselves, in a way that is completely determined by the properties of $H$ at the critical points. The number of negative eigenvalues of $H$, $k$, is  the \emph{index} of the critical point. The change undergone by the submanifolds $M_a$ as a critical level is passed is described using the concept of 'attaching handles'. Suppose that the critical level contains a critical point of index $k$. We define a $k$-handle $H^{n,k}$ in $n$ dimensions ($0\leq k\leq n$) as a product of two disks, one $k$-dimensional and the other $(n-k)$-dimensional:
\begin{equation}
H^{n,k}=D^k\times D^{n-k}.
\end{equation}
Having defined handles, we can state the main result of Morse theory. Let $\phi$ a smooth embedding $\phi:\mathbb{S}^k\times D^{n-k}\rightarrow\partial M$ ( $\mathbb{S}^k$ is a $k$-sphere). Then one can build the topological space $M\cup_{\phi}H^{n,k}$, i.e. $M$ with a $k$-handle attached by $\phi$. This procedure admits a generalization to the simultaneous attachment of $m$ $n$-dimensional handles $H_1^{n,k_1},\cdots,H_m^{n,k_m}$ of indexes $k_1,\cdots,k_m$.

\begin{figure}
\begin{center}
\includegraphics[width=0.2\textwidth]{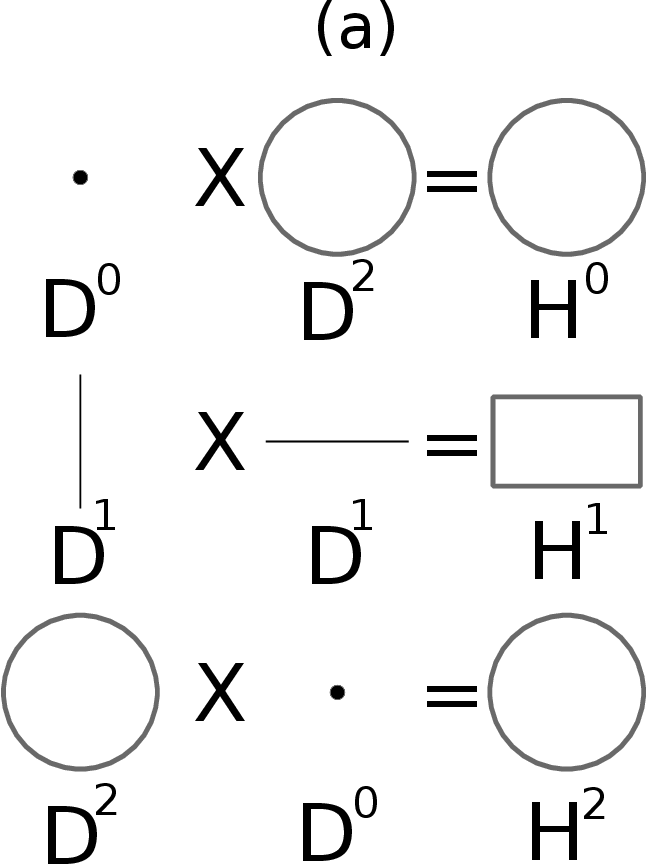}
\includegraphics[width=0.27\textwidth]{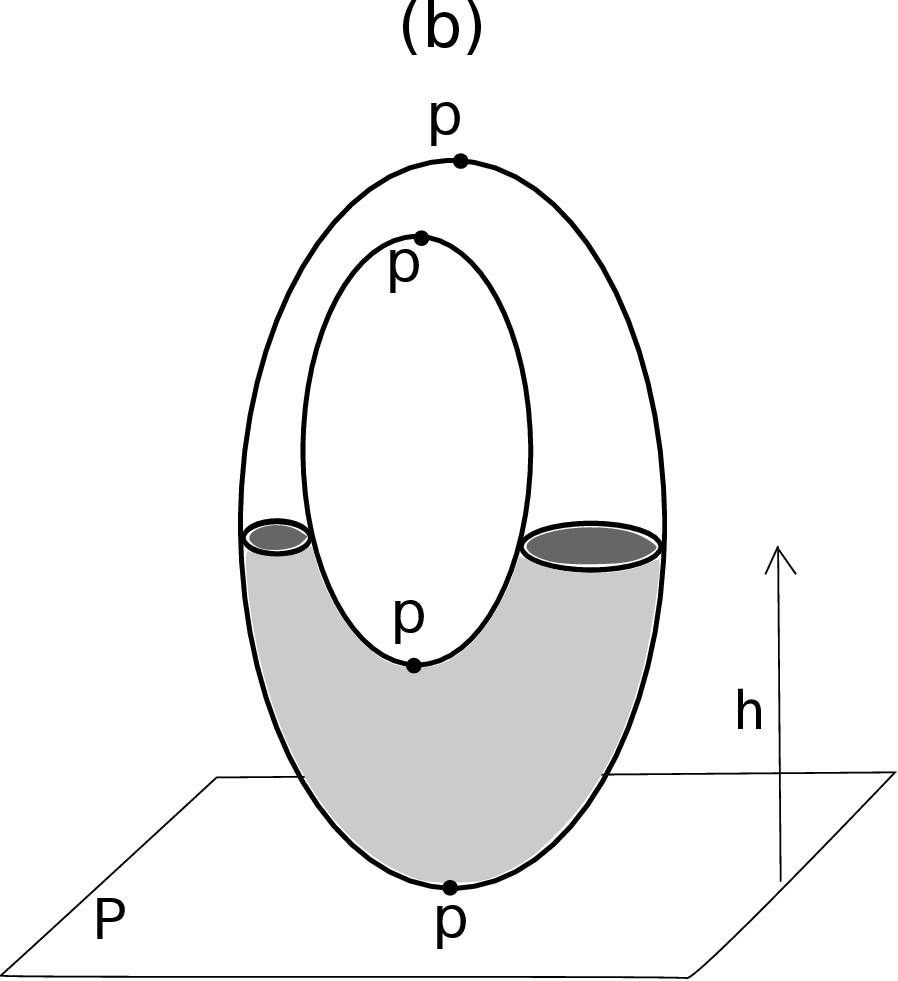}
\caption{(a) $2D$ handles; $H^0$ is the product of a $0$-disk and a $2$-disk, so that it is a $2$-disk. The same for the other two cases. (b) An example of a $2D$-torus. The height $h$ with respect to the plane $P$ is the Morse function, $M_h$ is the gray-colored part of the torus, $p_i$ for $i=0,1,2,3$ are the critical points of indexes $k=0,1,1,2$, respectively. The topology of the whole torus is reconstructed by attaching handles $H^0, H^1, H^1, H^2$, respectively at the critical levels corresponding with $p_i$ for $i=0,1,2,3$.}
\end{center}
\end{figure}

\section{Mean-field $|\phi|^3$ model}

Equation (\ref{char}) is equivalent to a $5$th degree equation in the magnetization $m$. Only numerical solutions are available for this equation. In order to check the results obtained here, we have introduced a slight modification in the m-f $\phi^4$ model to obtain an equation (\ref{char}) solvable analytically, i.e. of a degree less or equal to the $4$th. The goal has been get by lowering the degree of the local potential by one unit, i.e. by substituting the quartic term by a cubic term where the modulus has been added to conserve the $\mathbb{Z}_2$ symmetry, so that the local potential takes the form
\begin{equation}
V(\phi)=-\frac{\phi^2}{2}+\frac{|\phi|^3}{3}.
\label{vloc}
\end{equation}
$\nabla V=0$ becomes
\begin{equation}
\pm\phi^2_i-\phi_i-Jm=0,\quad i=1,\cdots,N,
\end{equation}
where the sign $+$ has to be taken if $\phi\geq 0$, while the $-$ if $\phi\leq 0$. The solutions are 
\begin{eqnarray}
\phi_1=\frac{1+\sqrt{A_+}}{2},
\\
\phi_2=\frac{-1+\sqrt{A_-}}{2},
\\
\phi_3=\frac{-1-\sqrt{A_-}}{2},
\end{eqnarray}
where $A_{\pm}=1\pm Jm$. By inserting in (\ref{char}) we get
\begin{equation}
am+b=c\sqrt{A_+}+d\sqrt{A_-},
\end{equation}
 where the coefficients $a=2N$, $b=N-2n_1$, $c=n_1$, and $d=-N+n_1+2n_2$ have been introduced. The last equation is equivalent to 
\begin{equation}
\left((am+b)^2-c^2A_+-b^2A_-\right)^2=4c^2d^2A_+A_-,
\end{equation}
which is a $4$th degree equation in $m$, q.e.d.. 

The results do not show any qualitative difference with respect to the m-f $\phi^4$ model, neither in the canonical thermodynamic, nor in the topology of the $v$-level sets. All the considerations made in Sec. \ref{phi4} for the m-f $\phi^4$ model can be perfectly transferred to the m-f $|\phi|^3$ model.

\begin{acknowledgments}
The most part of the work in this paper is contained in my Master's thesis \cite{baroni}, hence I warmly thank the supervisor L. Casetti.  Furthermore, I thank also M. Pettini for priceless discussion, suggestions, and for having introduced and continuously supported this line of research. 
\end{acknowledgments}


\begin{thebibliography}{99}

\bibitem{ising} E. Ising, \textit{Z. Phys.} \textbf{31} 253-258 (1925)
\bibitem{onsager} L. Onsager, \textit{Crystal Statistics. I. A Two-Dimensional Model with an Order-Disorder Transition}, \textit{Phys. Rev.} \textbf{65} 117 (1944)
\bibitem{goldenfeld} N. Goldenfeld \textit{Lectures on Phase Transitions and the Renormalization Group} (Cambridge: Perseus Publishing, 1992)
\bibitem{ccp} L. Casetti, E. G. D. Cohen, and M. Pettini \textit{Phys. Rev. Lett.} \textbf{82} 4160 (1999)
\bibitem{ccp1} L. Casetti, E. G. D. Cohen, and M. Pettini \textit{Phys. Rep.} \textbf{337} 237 (2000)
\bibitem{ccp2} L. Casetti, E. G. D. Cohen, and M. Pettini \textit{Phys. Rev.} E \textbf{65} 036112 (2002)
\bibitem{k} M. Kastner \textit{Rev. Mod. Phys.} \textbf{80} 167 (2008)
\bibitem{pettini} M. Pettini \textit{Geometry and Topology in Hamiltonian Dynamics and Statistical Mechanics} (Springer-Verlag New York Inc., 2007)
\bibitem{bc} F. Baroni, and L. Casetti \textit{J. Phys. A: Math. Gen.} \textbf{39} 529545 (2006)
\bibitem{b4} F. Baroni \textit{arXiv:}1903.125054v1 [cond-mat.stat-mech] (2019)
\bibitem{b} F. Baroni \textit{arXiv:}1102.3276v4 [cond-mat.stat-mech] (2011)
\bibitem{b1} F. Baroni \textit{J. Stat. Mech.} P08010 (2011)
\bibitem{b3} F. Baroni \textit{arXiv:}1611.09254v2 [cond-mat.stat-mech] (2016)
\bibitem{hk} I. Hahn, and M. Kastner, \textit{Phys. Rev. E} \textbf{72} 056134 (2005)
\bibitem{km} M. Kastner, and D. Mehta \textit{Phys. Rev. Lett.} \textbf{107} 160602 (2011)
\bibitem{mhk} D. Mehta, J. D. Hauenstein, and M. Kastner \textit{Phys. Rev. E} \textbf{85}, 061103 (2012)
\bibitem{dl} T. Dauxois, and S. Lepri \textit{report ENS-Lyon} (2001)
\bibitem{baroni} F. Baroni \textit{Master's thesis} Universit\`{a} di Firenze (2002)
\bibitem{aarz} A. Andronico, L. Angelani, G. Ruocco, and F. Zamponi \textit{Phys. Rev.} E \textbf{70} 041101 (2004)
\bibitem{bk} T. H. Berlin, and M. Kac \textit{Phys. Rev.} \textbf{86} 821-835 (1952)
\bibitem{cccp} L. Caiani, L. Casetti, C. Clementi, and M. Pettini \emph{Phys. Rev. Lett.} \textbf{79} 4361 (1997)
\bibitem{ck} L. Casetti, and M. Kastner \textit{Phys. Rev. Lett.} \textbf{97} 100602 (2006)
\bibitem{ck1} L. Casetti, and M. Kastner \textit{Physica A} \textbf{384} 318 (2007)
\bibitem{ckn} L. Casetti, M. Kastner, and R. Nerattini \textit{J. Stat. Mech.} P07036 (2009)
\bibitem{f} M. E. Fisher, The nature of critical points, in: W. E. Brittin (ed.), \emph{Lectures in Theoretical Physics} (University of Colorado Press, Boulder 1965), Vol. VII, Part c
\bibitem{fcsp} R. Franzosi, L. Casetti, L. Spinelli, and M. Pettini \textit{Phys. Rev.} E \textbf{60} R5009 (1999)
\bibitem{fp} R. Franzosi, and M. Pettini \textit{Phys. Rev. Lett.} \textbf{92} 060601 (2004)
\bibitem{fp1} R. Franzosi, and M. Pettini \textit{Nucl. Phys. B} \textbf{782} 219 (2007)
\bibitem{g} G. Gallavotti, \emph{Statistical Mechanics: A Short Treatise} (Springer, New York) (1999)
\bibitem{gfp} M. Gori, R. Franzosi, and M. Pettini, \textit{arXiv:}1602.01240 [cond-mat.stat-mech] (2016)
\bibitem{hk1} I. Hahn, and M. Kastner, \textit{Eur. Phys. J. B} \textbf{50} 311314 (2006)
\bibitem{huang} K. Huang \textit{Statistical Mechanics} (John Wiley and Sons, 1987)
\bibitem{ks} M. Kastner, and O. Schnetz \textit{Phys. Rev. Lett.} \textbf{100} 160601 (2008)
\bibitem{kss1} M. Kastner, S. Schreiber, and O. Schnetz \textit{Phys. Rev. Lett.} \textbf{99} 050601 (2007)
\bibitem{kss} M. Kastner, O. Schnetz, and S. Schreiber \textit{J. Stat. Mech.} P04025 (2008)
\bibitem{l} O. E. Lanford, "Entropy and equilibrium states in classical statistical mechanics," in \emph{Statistical Mechanics and Mathematical Problems}, edited by A. Lenard, Lectures Notes in Physics Vol. 20  (Springer, New York), 1 (1973)
\bibitem{lebowitz} J. L. Lebowitz \textit{Rev. Mod. Phys.} \textbf{71} S346 (1999)
\bibitem{yl} T. D. Lee, and C. N. Yang \textit{Phys. Rev.} \textbf{87} 404-409 (1952)
\bibitem{mk} D. Mehta, and M. Kastner \emph{arXiv:}1010.5335v1 [cond-mat.stat-mech]
\bibitem{palmer} R. G. Palmer \textit{Adv. Phys.} \textbf{31} 669 (1982)

\end{thebibliography}
\end{document}